\documentclass[aps,prmat,preprint,groupedaddress]{revtex4-2}

\usepackage{color}
\usepackage{lineno}
\usepackage{graphicx}
\usepackage[english]{babel}
\begin{document}

\preprint{}

\title[On the saturation stress of deformed metals]{On the saturation stress of deformed metals}

\author{S. Queyreau\textsuperscript{1}} \email{sylvain.queyreau@cnrs.fr}
\affiliation{\textsuperscript{1} Université Sorbonne Paris Nord, LSPM-CNRS, UPR 3407, 93430 Villetaneuse, France} 



\begin{abstract}
Crystalline materials exhibit an hysteresis behaviour when deformed cyclically. The origins of this tension-compression asymmetry have been fully understood only recently as being caused by an asymmetry in the junction strength and a reduced mean free path of dislocations inherited from previous deformation stage. Here, we investigate the saturation stress in fcc single- and poly-crystals using a Crystal Plasticity framework derived from dislocation dynamics simulations. In the absence of plastic localization and damage mechanism, the single-crystal mechanical response eventually saturates. We show that the cyclic saturation stress converges asymptotically to the monotonic saturation stress as the cycle plastic increment increases, and this convergence can be observed for some experimental conditions. The analysis of the experimental literature suggests that the mechanisms controlling the saturation in single crystals are the same controlling the cyclic response of polycrystals with large grains. We propose also analytical and approximated models to predict the saturation stress over the considered loading conditions. The saturation stress appears as a fundamental property of dislocations, explaining the consistency observed in the experimental literature. This work provides a unified view on the monotonous and cyclic responses of fcc single and poly-crystals, which may help in interpreting experimental data. 
\end{abstract}
\maketitle

\section{Introduction}

Deformation of crystalline materials depends upon the deformation history undergone by the materials. This is particularly apparent through the existence of hysteresis curves when alternating the loading direction in tension and compression as in cyclic deformation \cite{Bauschinger:1886, Brown:1971, Argon:2008}. A part of the deformation is reversible and the resulting hardening is much smaller that in continuous monotonic deformation, explaining why deformation can be easily accumulated this way. At every two cycles, the flow stress may increase usually in a continuous manner, until reaching a \emph{saturation shear stress} \cite{Cheng:1981, Mughrabi:1978,Lepisto:1986,Bretschneider:1997, Gong:1997,Li:2009, Li:2011}. In single crystals, this saturation stress depends upon the loading direction and the materials as it not always scales with the shear modulus \cite{Li:2011}. For a given loading direction and material, the saturation stress typically increases with the cycle shear increment $\gamma_{p,cy}$, and the saturation stress may well be below the stress values observed in monotonous deformation. For example, in Cu -certainly the most studied system- deformed in single glide condition, the cyclic saturation stress ranges from 16 to 40 MPa exhibits a three stage curve with a central plateau value of 28 MPa. In stable multislip conditions, the cyclic saturation stress typically increases monotonously with an apparent slope that depends upon the loading direction and the material \cite{Bretschneider:1997, Gong:1997,Li:2009, Li:2011} and may reach 50 or 60 MPa. In comparison the maximum stress reached by monotonically deformed single crystal of Cu is in the range of {80-100} MPa. 

Until recently, this tension-compression asymmetry -also known as the Bauschinger Effect (BE)- was commonly thought to be related to the building up of Long-Range Internal Stresses (LRIS) or backstress associated to the formation of dislocation patterns according to the so-called composite model \cite{Asaro:1975eu, Mughrabi:88, Argon:2008}. However, dislocation patterns are rather weak at smaller strain, they may well be different from the archetype of the composite model, and no LRIS was found in recent large scale mesoscale simulations \cite{Queyreau:2021} or by X-ray microdiffraction \cite{Kassner:2009fk, Kassner:2013fk} (except at very large strain \cite{Levine:2019}). Another explanation consisted in the partial dissolution of the microstructure \cite{Buckley:1956, Sleeswyk:1978, Depres:2008fk, Rauch:2011kx}, as a mechanistic way to reduce dislocation density and thus flow stress. However, no clear dislocation elementary mechanism was clearly identified as there is no in-situ observation of dislocation motion during cyclic deformation. Ultimately, these explanations explain neither the transient nature of the Bauschinger effect nor its reversibility component.

The present authors \cite{Queyreau:2020,Queyreau:2021} recently proposed a systematic study of cyclic deformations in single crystals by means of Discrete Dislocations Dynamic (DDD) simulations. Despite of the present of a pronounced Bauschinger effect in the simulations, no LRIS was measured in the simulations. Statistical analysis of the DDD simulations showed that the tension-compression asymmetry is caused by two original elementary mechanisms of dislocations. The first mechanism i) is related to easy destruction of binary junctions as their stability is asymmetrical as they formed from mobile segments whose curvature is driven by the applied loading. Junctions formed is tension are thus more stable in tension than in compression, and are easily destroyed during the backward motion of dislocation in compression. The second elementary mechanism ii) is related to the reduction of the mean free path of dislocations as they glide in regions of the crystal that they already explored. In the backward motion, the mobile segments unwind stored segments on the edge of the swept area, leading to a reduction in the storage rate. These mechanisms naturally explain the transient nature of the tension-compression asymmetry and the reversible component of plastic deformation. From these results, the authors proposed a modified Crystal Plasticity framework, that was implemented into a FEM to include the physics highlighted in the DDD. As a results this multiscale approach was successful in predicting the Bauschinger effect, hysteresis and saturation stress reported in most of the experimental literature on cyclically deformed fcc single crystals. 

Now that we have a CP framework justified from mesoscale physics and validated by confrontation with the existing single-crystal literature, we can now assess the full implications of these results and of the model. In particular, when comparing the literature data obtained though out decades on cyclic deformation of single crystals - the saturation stresses - observed in cyclic deformation are clearly more reproducible that the Bauschinger effect observed on a single tension-compression experiment. This fact suggests that the saturation stress in cyclic deformation is the sole result of the average of some basic dislocation mechanisms, while the Bauschinger effect is impacted  by the initial state of the material, such as the impurity content, the initial dislocation density or the microstructure, but also by the loading conditions. In other words, the saturation stress seems to be a basic property of the material and dislocations. Simulation results could also be processed in order to provide simple equations to help in interpreting and/or fitting the experimental data or to trace back to some of the fundamental constants of the CP model e.g. in an inverse approach. Finally, the single crystal is an integral part of the more complex polycrystal system. The elementary mechanisms described above are general enough to be operative in a polycrystal. The insertion of GB adds a new lengthscale, and a competition between this lengthscale and the ones associated with elementary mechanisms i and ii. These ideas will act as motivations of the present paper. 

The objectives of the present work are thus to extend and analyse the CP FEM results obtained for the cyclic deformation of fcc single-crystals. We will see that the saturation stress of cyclic loading is actually related to the theoretical saturation stress obtained in monotonous conditions. Analytical solutions to the constituting differential equations of the CP framework will be provided and can be employed to help in interpreting experimental data. Finally, the implication of the mechanisms i and ii on the cyclic deformation of polycrystal will be assessed.

\section {Methodology}

This section presents the Crystal Plasticity framework derived from the DDD results in \cite{Queyreau:2021}. Readers interested in the presentation of the technical details of the DDD technique along with a review of recent progresses are referred to \cite{Devincre:2011fk,Queyreau:BDM}. This CP model is an extension of the dislocation density based  descriptions proposed in the seminal work \cite{Kubin:2008fk, Kocks:03} and subsequent works inspired from DDD \cite{Devincre:2006uc} for monotonous loadings. In fcc metals at intermediate and high temperature, the flow stress is controlled by the formation and destruction of reactions -junctions- between dislocations belong to different slip systems. The critical shear stress $\tau_c^i$ on the active slip system 'i' is thus related to the \emph{forest} obstacle densities $\rho^j$ and expressed as :
\begin{equation}
\tau_c^i = \mu b \sqrt{ \sum_j a_{ij} \rho^j}
\label{eq:forest}
\end{equation}
Where $\mu$ is the isotropic shear modulus, $b$ the norm of the Burgers vector and $a_{ij}$ the components of the interaction matrix. These last interaction coefficients measure the average strength of the two interacting slip systems and can be determined in a straigthforward manner from DDD \cite{Devincre:2006uc, Queyreau:2009lr,queyreau:2010, Madec:2017kx}. For reasons of symmetry, only six different interactions of different nature exist among the twelve $a_0/2 <110>{111}$ slip systems existing in fcc metals. The interactions are the self-interaction, the coplanar interaction, three junctions: Lomer, Hirth and glissile reactions, and the collinear interaction. The interaction matrix has to have non identical components in order to reproduce fully the anisotropy of the plastic deformation of fcc single-crystals. 

A first extension of the CP framework  was introduced to capture the asymmetry of dislocation junctions (mechanism 'i' in the introduction) and to reproduce the tension-compression asymmetry observed in DDD \cite{Queyreau:2021}. Junctions formed during the forward loading from curved segments inherit a mechanical stability asymmetry. The junctions formed during prestrain or forward tension loading, are stronger when stressed in tension and statistically weaker when now stressed in compression, and this asymmetry in mechanical stability can be considered as a line tension effect, with the parent moving segments colliding when they were in extension. As a results, while plastic flow is still controlled by junctions destruction, the interaction coefficients during the backward loading are effectively weakened over a transient that depends on the amount of the initial forward deformation. The pool of weak junctions gets destroyed as  backward deformation proceeds, until mobile segments explore new area of the crystal and form junctions polarized in compression. A reversibility function $r_a$ was introduced to reproduce these effects as: 

\begin{equation}
a_{i j}^{b c k} = ( 1 - r_a ) \times a_{i j} \mathrm{, ~with ~~} r_a = \exp \left(- \frac{ \gamma_{b c k}^i} {C_a b \sqrt{ \Delta \rho_{pr}^i}} \right)
\label{eq:newa}
\end{equation}

with $a_{i j}$ the reference and constant interaction coefficient measured in continued tension. the subscripts 'pr' and 'bck' refer to prestrain and backward loading, respectively. In a cyclic loading, the prestrain  becomes the previous cycle, while the backward loading corresponds to the current cycle. The amount of backward strain impacted by this transient on the junction stability is approximated as $C_a b \sqrt{\Delta \rho_{pr}^i} $. This last equation can provide a lengthscale estimate to this effect. The density increase during previous deformation cycle is  ${ \Delta \rho_{pr}^i} = \max(\rho_{pr}^i) - \min (\rho_{pr}^i)$ and corresponds to the potentially impacted dislocation density polarized according to the initial loading. The constant $C_a$ has been determined as $C_a = 0.6 \pm 0.1$ through a statistical analysis of the interaction coefficients over a large panel of relevant DDD simulations \cite{Queyreau:2021}. The $\gamma_{b c k}^i / C_a b \sqrt{ \Delta \rho_{pr}^i}$  term within the exponential of $r_a$ states the competition between the easy destruction of junctions formed in tension, and the formation of new junctions in never-explored regions of the crystal. 

 
A second fundamental equation in the CP framework describes the evolution of the dislocation density on active slip system 'i' with the system shear $\gamma^i$. The evolution of the dislocation density is related to the kinetics of plastic activity that occurs through intermittent busts or avalanches of dislocation motion. In principle, dislocations are stored at the end of an avalanche as dislocation segments left at the edge of the swept area. This being said the exact theoretical connection between dislocation avalanches at the mesoscale and the observed continuous macroscale storage of dislocations has still to be formulated. For the present work and in monotonic loading conditions, the density evolution takes the following simple form: 

\begin{equation}
\frac{d \rho^i}{d \gamma^i} = \frac{1}{b} \left(  {1 \over L_{hkl}}  + {1 \over L_I} - y \rho^i \right) = \frac{1}{b} \left(  \frac{\sqrt{\sum_{i \neq j} a_{ij} \rho^j}}{K_{hkl}}  + \frac{\sqrt{a'_0 \rho^i}}{K_I} - y \rho^i \right)
\label{eq:KM}
\end{equation}
where the first two terms relate to the dislocation storage and the last term represents the dynamic recovery. The storage rate is commonly related to the Mean Free Path (MFP) of dislocations $L^i$, which represents the average distance covered by dislocations before the temporary or permanent storage. The MFP of dislocations typically depends upon the loading axis and thus the number of active slip systems. In \cite{Devincre:2006uc,Kubin:2008fk}, Kubin et al. proposed a simple decomposition of the MFP into elementary ingredients: i) the rate $p_0$ of forming a junction that will ultimately store a dislocation segment, and ii) the average length of stored dislocation $<l>$ and iii) junction segments and $<l_j>$, both of which scale with the dislocation density $<l> = k_0 / \sqrt{\bar{a} \rho}$ and  $<l_j> = \kappa <l> $. $\bar{a}$ is the average interaction coefficient for the considered orientation. Statistical analysis of DDD results showed that parameters $k_0, p_0, \kappa$ to be constants independent of the loading direction. The MFP can thus be written:
\begin{equation}
L^i = \frac{\mu b^2}{\tau_c^i} \left[ \frac{\sqrt{\bar{a}} n (1 + \kappa)^{3/2}}{p_0 k_0 ( n - 1 - \kappa )} \right]
\label{eq:MFP}
\end{equation}
the third term is associated to the dynamic recovery occurring at large dislocation density. $y$ is related to critical distance at which two dislocations can easily annihilate, which may be measured from atomistic data. Here, $y$ takes different values depending on the loading axis to reproduce the anisotropy of the onset of stage III observed on experimental deformation curves on single-crystal \cite{Takeuchi:1975, Kubin:2013fk}.

\begin{table*}[!ht]
\caption{List of physical parameters employed in the CP simulations of the cyclic deformation of fcc single crystals. Most of these parameters are coming from DDD simulations for monotonic or cyclic deformation, few parameters are coming from the experiemental litterature but for monotonic loading only. No parameters or equations were fitted to the cyclic deformation literature. The resulting FEM simulations are therefore truly predictive of cyclic deformation. }
\label{tab:paramCP}

\begin{center}
\small
\begin{tabular}{ c c c c c c c } 
\hline
 ${a'_0}$ (self) &   ${a_{ortho}}$ (Hirth)  & ${a_2}$ (glissile) & ${a_3}$ (Lomer) & ${a_{colli}}$ (Collinear) & ${\rho_{ref}}$ & ${\rho_{0}}$\\ 
 0.122 & 0.07 & 0.137 & 0.122 & 0.625  & $10^{12}\,$ m$^{-2}$ & $10^{12}/n_{sys}$  \\
 \hline
  ${K_{I}}$ & ${K_{112}}$ & ${K_{111}}$ & ${K_{001}}$ & ${C_a}$ & ${A_p}$ & ${C_p}$ \\ 
  90 & 10.42 & 7.29 & 5 & 0.6 $\pm$ 0.1 & 2 $\pm$ 0.6 & {2.3 $\pm$ 0.3} \\
 \hline
 ${\dot{\gamma}_{app}}$ & ${\dot{\gamma}^i_{0}}$ & m &  ${y_I}$ (SG, [112]) & ${y_{001}}$ (nm) & ${y_{111}}$ (nm) \\
 $n_{sys} \times {\dot{\gamma}^i_{0}}$ & 10$^{-4}$ s$^{-1}$ & 35 & 0.5 nm &  {3.6 (Ni), 3.4 (Cu)} & {2 (Ni), 1.5 (Cu) }\\
 \hline
\end{tabular}
\end{center}
\end{table*}

When considering the cyclic deformation, DDD simulations \cite{Queyreau:2020, Queyreau:2021} showed that dislocation evolution is still associated to the storage of segments in the wake of avalanches at the fringe of the area swept by dislocations, with some similarities to what was observed in alloys \cite{Queyreau:09}. Equation \ref{eq:KM} is thus still valid. However, the reduced storage (mechanism ii in the introduction) observed at the loading reversal where the stored segments are simply unwinded, needs to be taken into account through a modified MFP (through a change of rate of locking dislocation rate $p_0$). A second reversibility function $r_p$ is thus introduced \cite{Queyreau:2020, Queyreau:2021} as:
\begin{equation}
p_{0}^{b c k} = ( 1 - r_p ) \times p_{0} \mathrm{, ~with ~~} r_p = A_p \times \exp \left( - \frac{ \gamma_{b c k}^i} {C_p b \sqrt{ \Delta \rho_{pr}^i}} \right)
\label{eq:P0bck}
\end{equation}
Where $A_p$ and $C_p$ are two additional constants measuring the initial MFP drop and the transient length of the reduced MFP, respectively. Statistical analysis over the transient observed in DD showed that $A_a = 2 \pm 0.6$ and $C_p = 2.3 \pm 0.3$. Interestingly, the transient on the reduced MFP (ii) is much larger that the one on the reduced junction stability (i). These different transients explain the non-monotonic response at the beginning of the backward deformation that is sometimes observed on experimental deformation curves \cite{Ebener:1991, Daniel:1971, Nasu:1984}.

Finally, the flow rule provides a close form to the CP framework in connecting the plastic activity $\dot{\gamma^i}$ on the active system to the critical resolved shear stress $\tau_c^i$  and the applied shear stress $\tau^i$: 
\begin{equation}
\dot{\gamma^{i}} = \dot{\gamma_0^{i}} \left(\frac{\tau^{i}}{\tau_c^{i}} \right)^{m}
\label{eq:strainrate}
\end{equation}
In the case of fcc metals at room temperature, the  strain rate sensitivity is related to the formation and dragging of jogs along dislocation lines. Here, the constant $\gamma_0^i$ is taken as $\gamma_{app} / n_{act}$, with $\gamma_{app}$ the applied shear rate and $n_{act}$ the number of active slip system. The sensitivity exponent is chosen as $m > 35$ to stay as close as possible to the Schmid criterion. The proposed CP modeling focuses on the some fundamental mechanisms of dislocations that are described by some average constants obtained from statistical analysis of DDD results. The resulting FEM simulations are expected to be general and representative of fcc pure systems. The simulations  will not include plastic localization nor finite geometry effects for now. 

Great care has been paid to the numerical resolution of the set of ODE presented above using the Z-Set Finite Element software and Matlab. We employ a double nested Newton-Raphson implicit scheme to solve the corresponding non-linear equations and obtain $\rho^i$ and $\tau^i$ for a given time step. Gradients required in the Newton-Raphson are expressed analytically, and a relative convergence criterion was set to $10^{-7}$. The representative volume was set to a single point of integration in order to simulate several thousand of cycles of deformation and finite geometry effects are left for a future work. The parametrization and resolution strategy were validated in \cite{Queyreau:2021} by the one-to-one comparison of CP predictions with the reference DDD simulations. 







\section{Results}

\subsection{Saturation stress in monotonic deformation in absence of damage}

\begin{figure}
    \centering
    \includegraphics[width=1\textwidth]{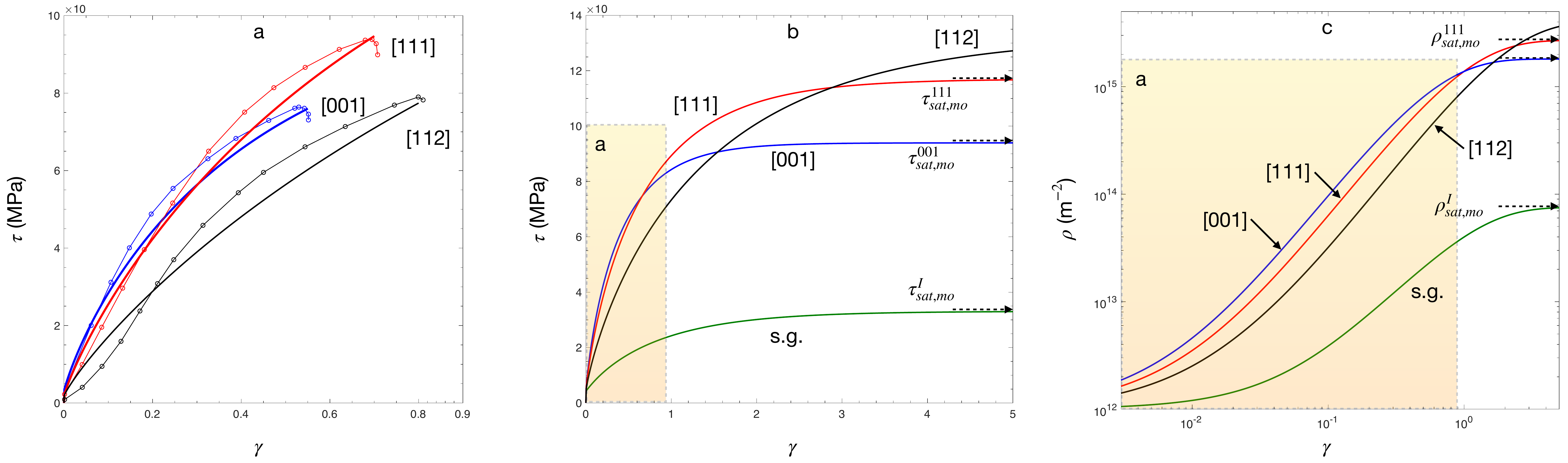}
    \caption{a. One-to-one comparison between the CP prediction (thick continuous lines) and the reference data from Takeuchi \cite{Takeuchi:1975} (circles) on the monotonous deformation of Cu single-crystals at RT. Key single-crystal orientations are chosen. b. Extension of the CP prediction until true saturation of the stress is obtained, in the absence of finite geometry and plastic localization effects. c. Corresponding total dislocation density evolutions in log-log scale. Note that the saturation density for [001] and [111] simulations are of the order of $> 10^{15}$ m$^{-2}$, which is still lower than the saturation density obtained in recent large scale MD simulations of single-crystals of Al \cite{zepeda:2021}.  
    }
    \label{fig:fig1}
\end{figure}

In this first section, we focus on the basic case of the single-crystal deformed monotonically in tension. This constitutes as well-pose problem, for which reference experimental data exists. The plastic response is mostly homogeneous during most of the deformation, in the sense that the slip activity is similar in all part of the single-crystalline sample \cite{Kahloun:2016}. We will see later that this basic plastic response can be connected to cyclic deformation.

First, to demonstrate the validity of the CP framework and its parametrization from DDD results, we start with a one-to-one confrontation of the CP prediction with some of the reference data from the experimental literature on single crystal deformation \cite{Takeuchi:1975}. Such experiments are rather delicate to perform as the materials state and experimental conditions can have a dramatic impact on the mechanical response, such as the impurity content, the precise orientation of the sample, or whether the jaws can rotate to accommodate crystalline rotation and ensure uniaxial deformation. Here, we compare our CP simulations with the reference work from Takeuchi on Cu single-crystals for selected loading directions: [112], [111] and [001] directions. The initial density was slightly adjusted in the simulations about $10^{12}$ m$^{-2}$ to capture the initial flow stress. 

This comparison is shown in Figure 1.a, where a nice quantitative agreement is found  between the simulations and the experimental reference data. The classical picture of the response of the single crystal is recovered here. The initial hardening rate depends upon the orientation and the number of slip systems activated simultaneously, e.g. in a increasing order: [112], [111] and [001] having two, three and four slip systems activated simultaneously. For [111] and [001] loading directions, these numbers of slips correspond to only half of the six or eight possible slip systems as these are pairs of colinear interactions. The colinear interaction is known to be a specific reaction among junctions, as it leads to the annihilation of the intersecting segments of dislocations and is associated to a very large hardening due to the shortening of mobile segments. DDD simulations have shown that when starting with pairs of colinear systems, one of the colinear system will take over the other for [111] or [001] simulations \cite{Devincre:2007}. Another notable aspect of the single-crystal response is the anisotropic onset of the stage III, when the dynamic recovery becomes noticeable. The dynamic recovery of the [111] in particular is weaker, so much so that the [111] curve eventually crosses the initially steepest [001] curve at a deformation of about 40\% of plastic strain. 

In agreement with the response of ductile materials, prior to the fracture of the sample, the experimental flow stress decreases after a maximum corresponding to the striction of the sample. This decrease is expectedly absent in the CP simulations as plastic localization and fracture mechanics are not included. Fracture mechanics may be included in different manner in FEM, for example using a cohesive zone framework \cite{Mishin:2010,Rice:1989}, in connection to atomistic mechanisms controlling the debonding of matter \cite{Vanderven:2004, Ehlers:2016,Ehlers:2017}. For now, let us focus on the plastic response in the absence of fracture mechanism. For a deformation corresponding to the experimental fracture, the hardening rate in the simulations has still a non-zero value. When recalling that  equation \ref{eq:KM} was proposed initially to saturate, the CP deformation curves are thus expected to saturate at a larger deformation. 

We thus expanded the deformation range until saturation of the stress is obtained, and this is shown in figure 1.b for Cu. All curves saturate eventually, the corresponding deformation at saturation is however rather large, even beyond 100\% for the [001] and [111] curves. The saturation stress in monotonous condition  $\tau_{sat,mo}$ follows the following hierarchy, with the single glide condition [135] having the smallest saturation stress of 28 MPa, [112], [001] and [111] having the largest $\tau_{sat,mo}$ = 140 MPa. Since $y_c$ is material independent for the [135] and [112] curves, the $\tau_{sat,mo}$ for these curves scales nicely with the shear modulus of the considered materials. This saturation stress values will be exploited a bit further. 

From a theoretical point of view, this saturation stress observed in the simulation can be predicted from the set of constituting ODE of the CP framework. The starting point is obviously the dislocation density evolution that has to be set to zero. In the case of single glide the resolution is straigthforward leading to: 
\begin{equation}
    \rho_{sat,I} = {\sqrt{a_0 '} \over K_I^2 y_{c,I}^2}
\end{equation}
The stable multislip conditions are a bit less straightforward to solve as the interaction coefficient exhibit a dependence upon the dislocation density inherited from the simplified line tension that is assumed in the critical stress equation \cite{Devincre:2008lr,queyreau:2010}. When the dislocation density is large this correction becomes important, which is the case at the large saturation stresses under consideration here. The logarithmic correction is defined as $c(\rho^i) = \ln \left( 1/ b\sqrt{a_{ij} \rho^j} \right) / \ln \left( 1/ b\sqrt{a_{ij} \rho_{ref}} \right)$, where $\rho_{ref}$ is the reference dislocation density used to determine the interaction coefficient $a_{ij}$. The saturation density is now:
\begin{equation}
    \rho_{sat,hkl} = \left( {\sqrt{a_0 '} \over K_I y_{hkl}} +  {\sqrt{ (n_{act}-1) \bar{a} c(\rho^i)  } \over K_{hkl} y_{hkl}}  \right)^2
\end{equation}
where $n_{act}$ is the number of activated systems. This equation is an implicit equation for $\rho^i$, but since the density is contained in a logarithm function, this equation converges after only few iterations to evaluate $\rho_{sat}$.  Finally, the saturation stress is obtained in inserting the $\rho_{sat}$ into the critical stress: 
\begin{equation}
    \tau_{sat} = \mu b \sqrt{\bar{a} n_{act}\rho_{sat}}
\end{equation}
In the past equations, the saturation stresses are solely function of interaction coefficients, MFP and $y$. In the model, these quantities are defined from statistical averages over evolving dislocation microstructures, and are weakly impacted by the nature of the fcc system under consideration (at least not as a  first order approximation).  The initial dislocation microstructure or density are absent from these equations. The saturation stresses appear thus as a \emph{fundamental dislocation property} representing the balance of storage and recovery processes among dislocations in fcc crystals. 

An approximate solution for the evolution $\rho^i (\gamma^i)$ is provided in the appendixes for single and stable multislip conditions. Theses functions can be used to interpret or fit experimental data. In monotonous condition, the dislocation density increases monotonically until saturation. The strain corresponding to the saturation density can be estimated from the density evolution as an integral: 
\begin{equation}
\Delta \gamma^i = \int_{\rho_0} ^{\rho_{sat}} \left[ \frac{1}{b} \left(  \frac{\sqrt{\sum a_{ij} \rho^j}}{K_{hkl}}  + \frac{\sqrt{a'_0 \rho^i}}{K_I} - y \rho^i \right) \right]^{-1} d\rho^i
\end{equation}
This only provides an estimation as the saturation density is only reached as $\gamma^i$ tends to $+\infty$. We now recover a dependence upon the initial dislocation density $\rho_0$ that affects the amount of deformation required before saturation. 


\subsection{Saturation stress in cyclic deformation in single crystals}

\begin{figure}
    \centering
    \includegraphics[width=1\textwidth]{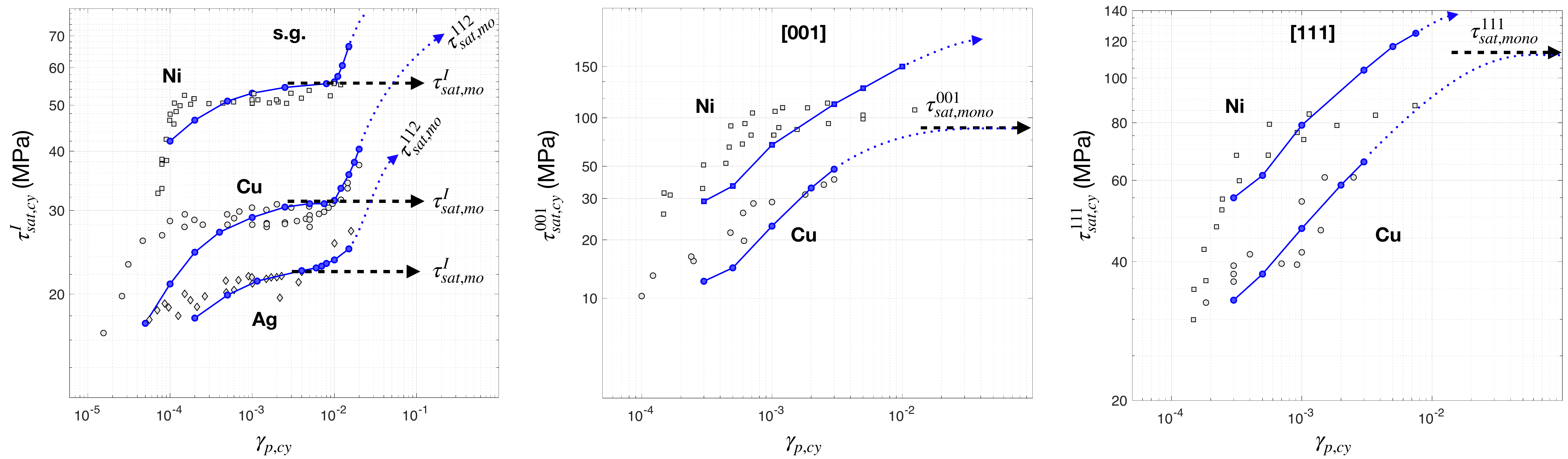}
    \caption{
    The saturation shear stress $\tau_{sat,cy}$ for cyclic deformation in fcc metals measured in a number of experimental studies \cite{Cheng:1981, Mughrabi:1978,Lepisto:1986,Gong:1997,Bretschneider:1997, Li:2009, Li:2011} compared to the prediction of the CP model in \cite{Queyreau:2020,Queyreau:2021} and to the theoretical saturation stress in monotonous conditions from previous section. Data is shown as function of the loading direction: a) initially single glide condition, b) [001] and c) [111] for Cu, Ni and Ag and the plastic strain increment $\gamma_{p,cy}$
    }
    \label{fig:fig2}
\end{figure}

The CP framework can be applied to alternating loadings and the simulations capture most of the features of cyclic deformation \cite{mughrabi:2010,Li:2011}. The maximum stress reached at each cycle increases monotonically until spontaneously saturating. For example in single glide conditions, we recover the so-called three stage curves of the now saturation stress in cyclic deformation $\tau_{sat,cy}$ with respect to the shear increment per cycle $\gamma_{p,cy}$. This is shown in figure 2.a for some fcc metals in comparison with most of the existing experimental literature \cite{Cheng:1981, Mughrabi:1978,Lepisto:1986,Gong:1997,Bretschneider:1997, Li:2009, Li:2011}. In the central plateau (or pseudo plateau) the microstructure typically transform from no Persistent Slip Band (PSB) to progressively fully constituted of PSB \cite{mughrabi:2010,Li:2011}. The plateau ends for strain increments of the order of 1\% when a secondary slip system activates. In the CP simulations, the center region of the curve is not exactly a plateau as it exhibits a slight slope. 

Figure 2.b and 2.c show the saturation stress for cyclic deformation of Cu and Ni single crystals oriented along [001] and [111] in stable multislip conditions. The saturation stress $\tau_{sat,cy}$ increases here monotonically without the presence of stages. The slope of the experimental curves seems however to decrease as the strain increment per cycle $\gamma_{p,cy}$ increases. This effect is more apparent on the data concerning Ni single crystals. The CP predictions of the $\tau_{sat,cy}$ are in nice qualitative and quantitative agreement with the existing experimental data. The CP predictions seem however to overestimate the saturation stress at larger  $\gamma_{p,cy}$, and this was suggested to be a consequence of the absence of dislocation microstructure in the CP simulations \cite{Queyreau:2021}. 

The saturation stress $\tau_{sat,cy}$ can be also obtained from the analytical resolution of the set of ODE presented in section 2. The resolution is however a bit more delicate with the presence of the two reversible functions $r_a$ and $r_p$ that are themselves function of $\gamma^i$ and the  dislocation density $\Delta \rho^i$ stored on the previous cycle. In cyclic deformation, we highlight that the saturation stress does not mecessary correspond to an horizontal tangent of the shear strain-shear stress curves of the considered cycle (except in the large $\gamma_{p,cy}$ limit discussed a bit further). At the end of a cycle at saturation, the hardening may well be non zero, but wince the cycle starts with a large dislocation density decrease, the increment over the entire cycle is null. To obtain the saturation density, one has to solve the integrated density evolution function: 
\begin{equation}
{\Delta \rho^i}=0= \int_{0}^{\gamma_{p,cy}/n_{act}} \frac{1}{b} \left(  \frac{\sqrt{\sum a_{ij} \rho^j}}{K_{hkl}}  + \frac{\sqrt{a'_0 \rho^i}}{K_I} - y \rho^i \right) {d \gamma^i} 
\end{equation}
Therefore, the dislocation evolution $\rho^i(\gamma^i)$ must be solved first contrary to the monotonous case. This can be done if the logarithm correction is approximated as a Taylor series where the dislocation $\rho^i$ in the current cycle is in the proximity of $\rho^i_{sat}$. The details are given in the Appendixes. The resolution leads once more to an implicit equation to obtain $\rho^i_{sat}$. 

Next, we connect the saturation stresses obtained in monotonous and cyclic deformation. Previous section has shown that the saturation stress in monotonous conditions can be obtained for rather large deformation amount. These saturation stress results could thus be drawn as well on the figure 2 for one cycle of the corresponding amount $\gamma_{p,cy}$. One would expect that the $\tau_{sat,mono}$ would act as an asymptotic limit to the $\tau_{sat,cy}$ curves. This is exactly what is obtained when considering the case of single glide condition. The saturation stresses obtained in single glide and monotonous conditions in previous section are strikingly close to the saturation stress at the plateau in cyclic conditions. The $\tau_{sat,cy}$ curves obtained from experiments or from the CP simulations asymptotically converge towards $\tau_{sat,mono}$. 

When now considering the third stage of the curves in figure 2.a or the multislip conditions in figure 2.b and 2.c, the agreement with experiments is a bit less obvious. The CP simulations are carried out past the range of $\gamma_{p,cy}$ considered in the experiments. Similarly to the single glide condition, the simulated $\tau_{sat,cy}$ curves converges towards the $\tau_{sat,mono}$ limit defined in previous section. In figure 2.a, the third stage converges certainly toward the $\tau_{sat,mono}$ in secondary double glide, that is probably out of reach. The agreement with the experimental data is rather good for the case of Cu. However, the quantitative agreement between with the model is a bit less good for Ni single crystal data, nonetheless the experimental data seem to enter a plateau for the largest $\gamma_{p,cy}$ considered experimentally, which could be well agree with the analysis proposed here, but not the quantitative values of the saturation stress. The $\tau_{sat,mono}$ being here overestimated in the model due maybe because of some specific parameters less well documented for Ni, such as $y_{hkl}$ or the stronger impact of microstructures that are absent in the model.  


\subsection{Saturation stress in polycrystals}

\begin{figure}
    \centering
    \includegraphics[width=0.5\textwidth]{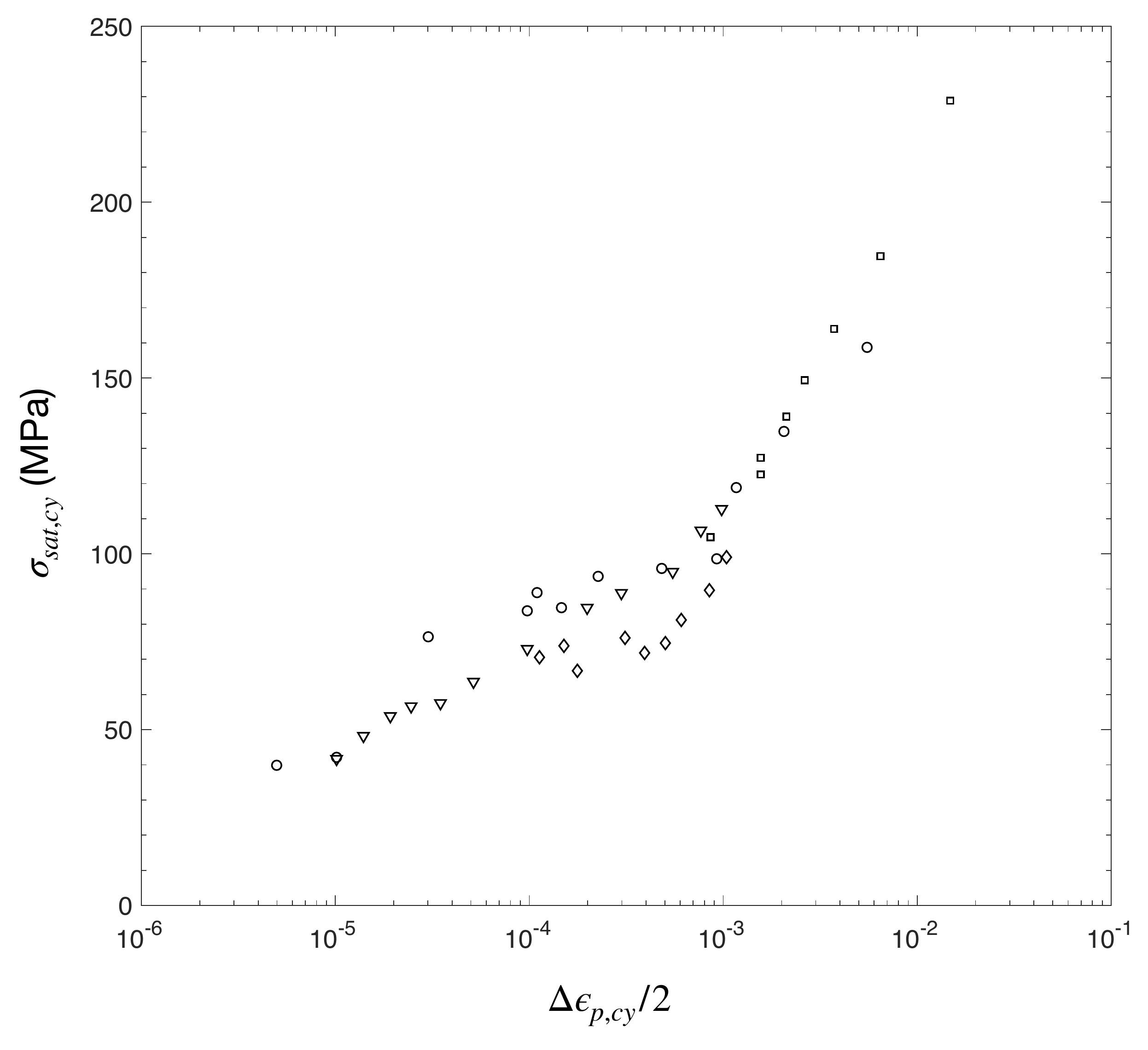}
    \caption{
    compilation of saturation stresses obtained experimentally on polycrystalline Cu with large grain size \cite{Magnin:1984} as function of strain increment $\Delta \epsilon_{p,cy}$. The data covers the work from Saxena and Antolovich (1975, squares), Figueroa et al. (1981, circles), Mughrabi and Wang (1981, triangles) and Rasmussen and Pederson (1980, diamonds)
    }
    \label{fig:fig3}
\end{figure}

This last section aims at connecting our results on the single crystal to the more complex and general polycrystalline system. What comes next is simply an illustration as the plastic response of the polycrystals is the average of many different features or elementary mechanisms that are not addressed here, ranging from the crystallographic texture, grain size and morphology, and dislocation-grain boundary interactions. Besides, simulating using CP FEM the cyclic deformation over more than 10,000 cycles while accounting for the a representative polycrystalline microstructure seems computationally out of reach, still to date. This being said we reprise a part of the analysis initiated by Magnin et al. in \cite{Magnin:1984}, where they compiled the cyclic response of polycrystalline Cu with relatively large grains. The collection of saturation stresses $\sigma_{sat}$ from the literature is displayed in figure 3. These curves typically show an increase in the $\sigma_{sat}$ with the plastic increment per cycle $\epsilon_{p,cy}$, some curves may suggest a quasi plateau regime starting for $\epsilon_{p,cy} > 10^{-5}$ and ending around  $\epsilon_{p,cy} \approx 10^{-3}$. The deformation microstructures transform from veines to PSB that are typical of the patterns observed in single glide conditions in single crystals. For $\epsilon_{p,cy} >\approx 10^{-3}$, the saturation stress increases more rapidly. The deformation microstructures now correspond to mazes, that are typical of the patterns observed in single-crystals deformed in multislip conditions. 

The polycrystalline plastic response can be understood in terms of effective single-crystal response through the so-called Taylor coefficient $M$, which depends upon the crystallographic texture. The so-called Hall-Petch effect is neglected and this is a valid assumption in the case of large grain polycrystals. In the absence of precise $M$ measures, the Taylor or Sachs hypothesises are commonly considered as bounding limits. The figure 4 shows theses two transformations of the polycrystal saturation stress into a single crystal approximation as $\tau_{sat} = \sigma_{sat} /M$. The Sachs is typically a good approximation at small strain, while the Taylor approximation works well at large strain. The average single crystal behaviour is expected to follow Sacks approximation first and transition to the Taylor ones at larger deformation. These curves will be compared with the saturation stress obtained experimentally on single crystals. At small strain, one might expect that grains in the polycrystal deform in single glide condition as suggested by the dislocation microstructures, and we see a sticking agreement between the single crystal $\tau_{sat}$ and the Sachs approximation for $\epsilon_{p,cy} < 5.10^{-4}$. For very small strain, so grains may remain in the elastic regime. For large deformation, one might expect that all grains to deform in multislip conditions, in accordance with the maze microstructure observed experimentally, and the [001] curve in particular remains between the two approximations. Between these two extreme situations, the polycrystalline behaviour is certainly a composite response of the single glide and multislip grain activity, with more and more grains transitioning into multislip as $\gamma_{p,cy}$ is increased. 

\begin{figure}
    \centering
    \includegraphics[width=0.5\textwidth]{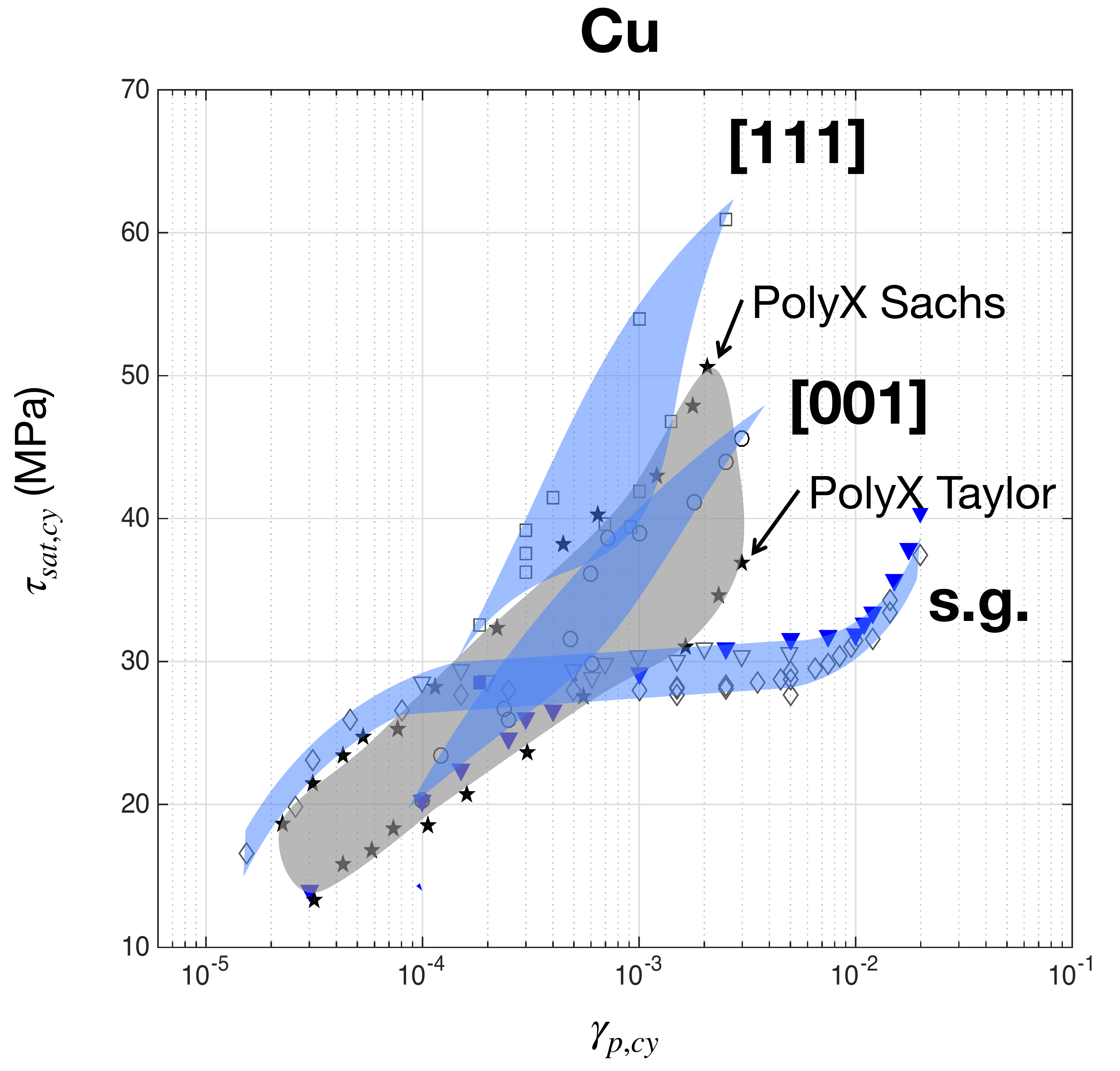}
    \caption{
    Comparison of the cyclic saturation shear stress of single crystals and polycrystals in pure Cu from Mughrabi \cite{Mughrabi:1981} and Wang  using the Taylor coefficient following the procedure from Magnin et al. \cite{Magnin:1984}. 
    }
    \label{fig:fig4}
\end{figure}

These correlations may thus explain the shape of the raw $\sigma_{sat}$ curves from the previous figures. Most grains of the polycrystals are thus following the single crystal response, which will eventually reach a saturation stress corresponding to $\tau_{sat,mono}$ of 28 MPa. The plateau in the single-crystal system is transformed in a quasi-plateau in the polycrystalline system as more grains transition into multislip conditions.  At large $\epsilon_{p,cy} =\epsilon_{III}  \approx 10^{-3}$ the polycrystal enters stage III where most of the grains are deformed in multislip conditions. Interestingly, the curves for single slip and [001] obtained for the single crystal cross for a deformation $ \approx M \gamma_{III} = M*3e{-3}$ that corresponds well to the onset of the third stage $\epsilon_{III}$ observed in the figure 4. The cyclic behaviour of the polycrystal with large grains seems thus to correspond simply to the average of the individual single grain behaviour, similar to what is accepted for polycrystals in monotonic deformations. 

Finally going back to our CP framework, we showed that its predictions were in quantitative agreement with the single crystal cyclic response, and we proposed expressions to predict the saturation stress in these conditions. The previous qualitative analysis means that we can use these results to estimate the saturation stress of the polycrystal with large grains as well.


\section{Conclusion}

In this paper, we employ a physically based CP FE model derived from our recent DDD analysis \cite{Queyreau:2020,Queyreau:2021} to analyse the saturation stresses obtained in monotonic and cyclic deformations of single-crystals and polycrystals of fcc metals.  

\begin{itemize}
    \item First, we compare the model results to reference strain-stress deformation curves from the experimental literature on Cu single-crystals. In the absence of plastic localization and fracture mechanism, the CP model expectedly saturates at large strain. 
    \item For cyclic deformations, the CP model reproduces the cyclic saturation over various loading conditions and fcc metals. The saturation stress observed in monotonic conditions acts as an asymptotic behaviour at large cycle increment strain $\gamma_{p,cy}$ to the cyclic deformation. This asymptotic behaviour is clearly reached in the plateau stage of single-glide conditions for several fcc metals and for Ni single crystals in multi-slip conditions at large strain increment. 
    \item From an analysis of the experimental literature on large grain polycrystals, we have shown that the polycrystalline response to cyclic deformation corresponds to the response of an aggregate of effective single-crystals. At small strain increment, grains are deformed elastically or in single slip, a quasi plateau can thus be sometimes seen as in single-crystals. Then at larger strain, grains transition into multislip conditions with larger saturation stresses.
    \item We proposed analytical or approximated solutions of the CP ODE to predict the saturation stresses in single or multislip conditions for single crystals. These models can be employed to interpret experimental data. 
    \item These results were obtained in the absence of plastic localisation and dislocation microstructure, which means that these features have only a secondary order impact on the macroscopic behaviour. 
    \item The saturation stress in monotonic or cyclic conditions appear as a fundamental property of dislocations mechanism as it is simply related to averages of dislocations interactions (through the interaction coefficients), dynamical recovery mechanism (through $y$), and reversibility part of the microstructure. 
    \item These quantities are phenomenological averages from DDD, where large number of binary interactions are occurring in absence of patterns. 
    \item Experimental data on the saturation stress can thus now be used to define these physical quantities and the procedure can be applied to other bcc or hcp crystalline systems.
\end{itemize}

\bibliography{bib_sat_stress_metals.bib}

\end{document}